# Intelligent Interface Architectures for Folksonomy Driven Structure Network

Massimiliano Dal Mas

*me @ maxdalmas.com*

*Abstract* — The folksonomy is the result of free personal information or assignment of tags to an object (determined by the URI) in order to find them. The practice of tagging is done in a collective environment. Folksonomies are self constructed, based on co-occurrence of definitions, rather than a hierarchical structure of the data. The downside of this was that a few sites and applications are able to successfully exploit the sharing of bookmarks. The need for tools that are able to resolve the ambiguity of the definitions is becoming urgent as the need of simple instruments for their visualization, editing and exploitation in web applications still hinders their diffusion and wide adoption. An intelligent interactive interface design for folksonomies should consider the contextual design and inquiry based on a concurrent interaction for a perceptual user interfaces. To represent folksonomies a new concept structure called "Folksodriven" is used in this paper. While it is presented the Folksodriven Structure Network (FSN) to resolve the ambiguity of definitions of folksonomy tags suggestions for the user. On this base a Human-Computer Interactive (HCI) systems is developed for the visualization, navigation, updating and maintenance of folksonomies Knowledge Bases – the FSN – through the web. System functionalities as well as its internal architecture will be introduced.

## I. INTRODUCTION

Both the scientific communities and the industry recognized in the last years the role of folksonomy for knowledge management, access and exchange.

Generally users cannot be assumed to have formal skills that are required nowadays to interact with folksonomy Knowledge Bases (KBs) with respect to both the contribution to the KBs' growth (as its updates and maintenance) and the access to the KB's content (through query and navigation).

On the paper [1] was proposed the use of folksonomies and network theory to devise a new concept: a "Folksodriven Structure Network" to represent folksonomies. The network structure of Folksodriven tags – *Folksodriven Structure Network (FSN)* – was thought as a "Folksonomy tags suggestions" for the user on a dataset built on chosen websites [2, 3]. A *FSN* can be considered as a way to solve the Ontology Matching problem between Folksodriven tags – that are hard to categorize [4].

Analyzing *FSN* it was showed as it can facilitate serendipitous discovery of content among users.

A *FSN* determines the flow of information in folksonomy tags network and determines its functional and computational properties [2, 3]. Previous works have addressed in the development of an interface for visualizing and navigating tag structures [5, 6, 7, 8, 9, 10, 11]. Currently an arduous matter could be on developing a collaborative environment (Web 2.0) according to the users' perceptions on dynamic systems based on folksonomy KBs – represented by the *FSN* – in the updating, and the retrieval information steps. The Semantic Web (Web 3.0) can help in that.

In this paper is briefly introduced a system to navigate, query and update folksonomy KBs. The proposed solution allows exploring the concepts and their relational dependencies as well as the instances by means of hyper-links; moreover, it provides a front-end to query the repository with SPARQL query language. The prototype has mainly been tested in the news websites domain.

## II. FOLKSODRIVEN NOTATION

Folksonomies are trying to bypass the process of creating ontologies supporting the development of evolutionary terms (and relations within a subject area). Folksonomies involve well-known problems and defects, such as the ambiguity of the terms and use of synonyms. Skeptics think that the folksonomy represents the first step toward anarchy of the Web but the Web is already an anarchist, which are both its greatest strength and its greatest weakness. As the Web grows, the best way to generate the metadata for the resources of the Web will always be a central issue.

Folksonomies represent exactly one of the attempts in place to monitor and categorize Web content to limit the costs to standstill "anarchy of the Internet".

The network structure of "Folksodriven tags" – *Folksodriven Structure Network (FSN)* – was thought as a "Folksonomy tags suggestions" for the user on a dataset built on chosen websites.

$$(1) \quad FD := (C, E, R, X)$$

As stated in [1, 2] we consider a Folksodriven tag as a tuple (1) defined by finite sets composed by:

• Formal Context (C) is a triple C:=(T, D, I) where the objects T and the attributes D are sets of data and I is a relation between T and D [4];

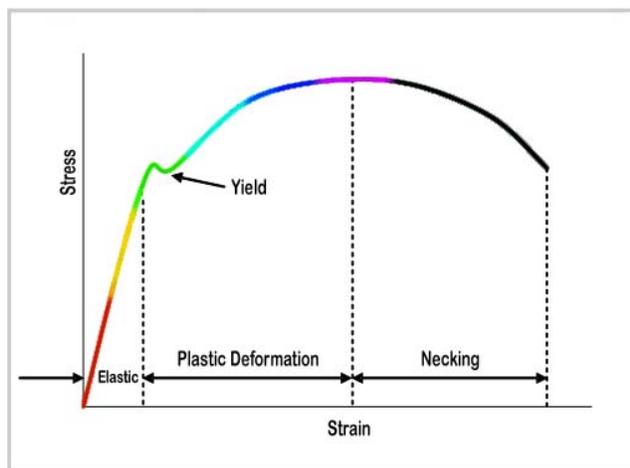

Figure 1. Stress-strain curve

• Time Exposition (E) is the clickthrough rate (CTR) as the number of clicks on a Resource (R) divided by the number of times that the Resource (R) is displayed (impressions);

• Resource (R) is represented by the uri of the webpage that the user wants to correlate to a chosen tag;

• X is defined by the relation $X = C \times E \times R$ in a Minkowski vector space [12] delimited by the vectors C, E and R.

### III. FOLKSODRIVEN STRUCTURE NETWORK

A *Folksodriven Structure Network (FSN)* consists of folksonomy tags (e.g. collections of tags about news websites) which are arranged regularly in Minkowsky vector space defined by *Formal Context (C), Time Exposition (E)* and *Resource (R)* as defined in [1].

The arrangement is a repetition of the smallest unit about a subject (e.g.: news websites regarding football match, airplane traffic, etc.), called a unit cell, resulting in the periodicity of a complete *FSN*. The *FSN* lattice represents a network that correlate different *Folksodriven tags (FD tags)* and so different folksonomy tags [1] chose by a user on a group of websites (here we consider a group of new websites). So a *FSN* lattice performs an Ontology Matching between different and unordered Folksodriven tags – that are hard to categorize.

A direct impact on network connectivity (structural plasticity) of the sources considered (i.e. the website considered) could be done by Morphological changes, such as changes in folksonomy tags chose by the user. As a consequence of morphological changes, links between network tags may break and new links can be generated. Local structural changes at the single *FD tag* may entail alterations in global network topology of the *FSN*. Conversely, global topology can have impact on local chose on *FD tag* by the user on a new web source to correlate with a folksonomy tag.

The stress σe could be plotted against the strain εe to measure the ontological plasticity to obtain an engineering stress-strain curve such as that shown in Figure 1. The curve is drawn with a rainbow grade color from the red to indicate the elastic part and the green for the yield part, to the purple-black for the necking part. The colors used to depict stress–strain values are used for the Information Architecture as "metaphor" to help the user on the navigation as illustrated in the next paragraphs.

### IV. INFORMATION ARCHITECTURE

Nowadays we use the Web to provide vast quantities of information and to communicate with the users. To wrest with today's complex information challenges we could have a wide variety of options to approach to the problem. Different approaches provide different kinds of ways of seeing and find the right path for a web user.

Web information design or Information Architecture (IA) works to define and optimize a systematic approach for the design method of online information spaces [13]. A well developed IA should provide users to envisage quickly where they are and where they can go, let them to perceive the necessary steps to reach their target [14, 15, 16].

For the basic requirements of IA: information must be easily retrieved by the users, who may have no prior knowledge of how it is organized. Information need to be collected using a metaphor, labeled and organized according to a navigation design.

#### A. Metaphor

A metaphor on a web site (i.e.: a bread-crumb, a path, a tree, etc.) can give to the users the initial clue to look for relevant elements as they travel through the site. It should be clarified when a graphic is just an illustrative ornament and when it is functioning as a signpost.

#### B. Labelling

Labeling make it clear what is behind each link. They could all be grouped into one section to be explained clearly. In this way with clear labeling information can result valid to the users.

#### C. Navigation

The navigation design targets are fast and easy for location of information within the site, ensuring that the user knows where they are and also where they should go next from any point on the site. Navigation design combines site structure with clearly labeled links and effective search tools to have a usable website.

To support navigation, query, and editing of ontologies for users with little or no knowledge of formal languages in which they are represented, a number of functionalities should be implemented.

## V. NAVIGATION DESIGN

Navigation design towards *FD tags* seems like a simple business, but if we scratch the surface the complexities of navigation design become merely apparent.

The navigation design is useful for a cohesive system that is a bare-bone depiction of all the *FD tags* that compose the *FSN* and how they fit together. The navigation becomes the place where IA and visual design come together.

We still don't know just how (or how much) people keep a folksonomy structure in their heads: the best approach is to assume that users carry no knowledge on those. A navigation system enables the user to navigate the *FD tags* successfully in a variety of circumstances.

The *FD* navigation design must accomplish three simultaneous goals:

- first it must provide users with a means for getting form one *FD tag* to another *FD tag*;
- the *FD* navigation design must communicate the relationship between the elements contained. It's not enough to merely provide a list of links. What do those *FD tags* have to do with each other? Are some more important that others? What are the relevant differences between them? This communication is necessary for users to understand what choices are available to them;
- the *FD* navigation must communicate the relationship between the proposed *FD tags* and what the user is currently viewing. Communicating this helps users understand which of the available choices might best support the task or goal they are pursuing.

## VI. FOLKSODRIVEN ONTOLOGY NAVIGATION

The ontology Knowledge Base (KB) for the *FD* structure is based on Description Logics (DL) [16] composed by the T-Box (terminological box) and the A-Box (assertional box).

On T-Box statements are described with classes in object-oriented languages. The T-Box contains sentences describing concept hierarchies (i.e. relations between concepts).

While A-Box statements are described with instances of T-Box classes. The A-Box contains "base" statements describing specific individuals and indicating where in the hierarchy they are positioned (i.e. relations between individuals and concepts).

Therefore T-Box statements tend to be more stable and permanent than A-Box statements.

To perform ontology navigation, both T-Box and A-Box statements should visualize the structure defined by their statements [17].

The user should be supported not only for a navigation of concept hierarchies defined by *isA* relations, but also on other forms of collection of metaphor and labeling used to organize the navigation as:

- locations or parts of a classification can be linked through a *PartOf* relation and it should be possible to group a set of sub-classification "under" the one they are all subparts of (e.g. to browse all sub-classification of a classification we can use "*isComposedOf*");
- the time relation to sort individual members from the first to the last one is exploited using an historical periods ordered according to a relations such as *FollowedBy*.

While T-box editing maintenance requires a skilful knowledge about ontological formalisms, A-Box editing should be supported, to allow the addition of new individuals to the KB and to allow the maintenance of the KB.

To develop this functionality we have to take into account the following points:

- range and cardinality restrictions defined in the T-Box need to be respected;
- contextual editing of the A-box while querying/maintaining the ontology, should be supported.

We consider the development of the interfaces enabling non expert users querying the ontology (e.g. visual query forms).

In the following paragraphs, we discuss an environment for navigating, querying and editing the A-Box of Web Ontology Language 2 (OWL2) ontologies (up to OWL-DL) through a web-based interface.

*1) Navigation*

The users can visualize ontology individuals and their properties – browsing properties via hyperlinks – through the ontology navigation interface. Browsing ontologies is possible to explore the available information and to refine the user search requirements – even when starting search with no specific requirement in mind [18].

An "interactive pie chart" is used to graphically represent the hierarchical organization of the different concepts and individuals of the ontology.

Charts represent an excellent way of visualizing data. Usually charts are created from tables or spreadsheets, which require significant manual data entry by the user. A more intuitive and easy way to edit and update data could be an interactive pie chart.

This work describes the ontology construct used for the folksonomies chosen by the users respect a graphical interaction of a pie chart representation. The graphical HCI idea developed here can also be applied to different kinds of charts (eg.: bar charts, line graphs, etc.). For our implementation the users can edit the pie chart towards a direct manipulation of the chart using the mouse.

The aim of the navigation pie chart is to support the exploration of the *FD* ontology, to support the visualization of classes and instances highlighting the relationships between them.

*Folksodriven tags (FD tags)* are used to make a clear *labeling* on pieces of information that could be grouped into different classes. Ontological classes are so represented by *FD tags* in different sectors of the pie chart using different colors according to their *stress-strain* (see *Folksodriven Notation* and [3]), to give to the users the initial clue using the principle of *metaphor*.

The relations between *FD tags* are done by the *Folksodriven Structure Network* (*FSN*) that could be modified by the users by the pie-chart interface (see *Folksodriven Structure Network* and [3]).

Information is so collected using a *metaphor* (the pie chart), *labeling* (the *FD tags*) and organized according to a *navigation* design (*FSN* visualized with the colored pie chart).

The pie chart does not only represent the hierarchy of classes connected through *isA* binary relations (like a navigation tree of Protégé), but it also represents the connections among individuals identified thanks to domain dependent classes of properties (e.g. *PartOf*, *isLocatedIn*, and so on). Ontological relations supporting visualization for the dynamic pie chart must be not symmetric and their inverse must be functional, consequently these criterions the *FD* ontology structure is identified by directed acyclic graphs.

According to these properties an individual is directly linked to its "father". This is particularly relevant with respect to relations defining hierarchical spatial and temporal structures (e.g. representing historically relations or restraint relations) to "mereological" relations (e.g. *PartOf*, *composedOf*).

For the pie chart visualization we depict relations defining total orders on individuals (e.g. *isFollowedBy*).

The base for the navigation pie chart is the OWL class *Thing*, and the rest of the pie chart division is organized as follows (see Figures 2-3-4):
• under the main pie chart sector, there are the top-level classes (i.e. direct subclasses of *Thing*);
• each class can be expanded to show its subclass hierarchy and its individual members visualizing the sub-sectors (as in Figure 3);
• individual-to-individual pie chart sector connections are defined according to a number of selected part-of properties (e.g. *PartOf* );
• finally, if total order relations is selected (as in Figure 4), they are exploited to order individuals within a given level (e.g. *isFollowedBy*)

An example of a navigation pie chart is presented in Figure 2. In this example *FD tag* "sinking" is a class and ship, captain, rescue, etc... its relative instances. In Figure 3 the FD tag "passenger" is another class and ship, plane, train, etc... its relative instances. The single individuals – for both Figure 1 and 2, in turn, are connected each other by the *PartOf* directly property (see Figure 3).

While in Figure 4 the user combined both *FD tags* "passenger" and "sinking" defining the total orders on both individuals by *isFollowedBy* that is another class and we obtain ship, ferry, titanic, etc... as its relative instances. The single individuals, in turn, are connected each other by the *PartOf* directly property.

The colours used in the diagrams represent stress–strain values for the elasticity in the *FD* ontology matching for the *FD tags* used as stated in "Folksodriven Structure Network paragraph" and in [3], see Figure 5. The stress-strain is a relationship between stress, derived from measuring the load for time exposition (E) applied on the lattice of the *Folksodriven Structure Network (FSN)*, and strain, derived from measuring the deformation of the *FSN* for the ductile range sets on *formal context (C)* related to the *time exposition (E)* – i.e. elongation, compression, or distortion of the *FSN*.

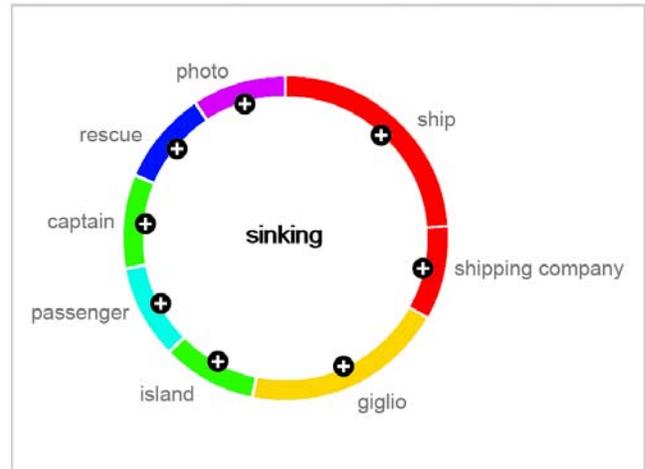

Figure 2. Navigation pie chart for the *FD tag* "sinking", the single individuals are connected each other by the **partOf** directly property

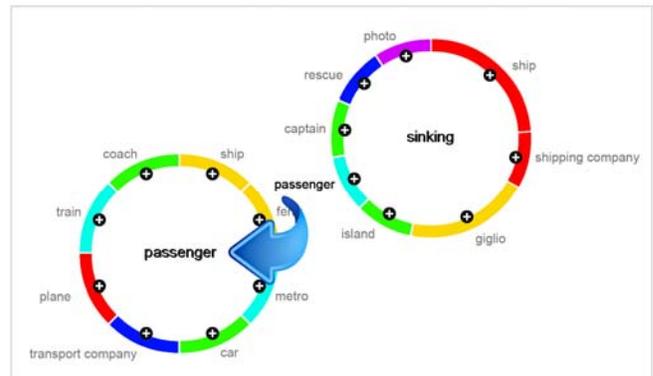

Figure 3. Navigation pie chart for the *FD tag* "passenger" derived by the above *FD tag* "sinking ", the single individuals are connected each other by the **partOf** directly property

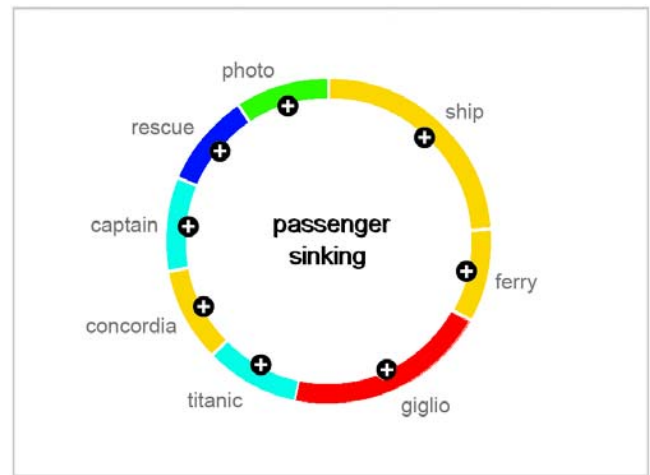

Figure 4. Navigation pie chart for thecombined *FD tags* "sinking" and "passenger" defining the total orders on both individuals by *isFollowedBy*

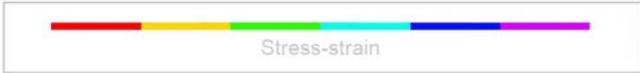

Figure 5. *Stress–strain* color rapresentation for the elasticity in the *FD* ontology matching

*2) Editing*

The application allows users creating, editing and removing individuals of the constructing ontology, their properties and also their labels. To ensure a multi-classification support, it is possible to define several labels for every individual.

Users can also create new individuals related to an existing one by means of selected part-of properties (e.g. *PartOf* ) (as shown in Figure 3): the new individuals are immediately displayed on a pie chart defined from the selected "father".

The properties of each class are defined in the T-Box. Two types of properties are distinguished: object property is a binary relation between two individuals and datatype property is a binary relation between an individual and a literal (a primitive type, like string or number).

Cardinality and range restrictions for properties are used to support users while editing. For example, in news *FD* ontology, the class *TypologyOfNewsObject* has the property *builtOf* . This property has no cardinality restriction (so it can have zero, one ore more values) but Ship is specified as range (co-domain). For instance as depicted in Figure 6, Sinking is an instance of *TypologyOfNewsObject* and has the property *builtOf* Passenger, where Passenger is an instance of Ship, see the result on pie chart navigation on Figure 2 and 3.

There are data types and object editor for the user contribution to the *FD tags*. The data type editor allows editing new literal values towards a text input box, while object allows selecting the relative property values (see Figure 7) that will be presented in the pie chart navigation. In Figure 7 we see the creation of a new *instanceOf* for the literal value "New Value". The properties presented in the pie chart are only those that are valid for the property range (see Figure 6).

*3) Querying*

The query interface was implemented using SPARQL. In this interface users can directly write queries expressed in the SPARQL language [19], display results in pie chart form and navigate through results via hyperlinks. This interface is very flexible because users can write arbitrary queries but suitable only for expert end users.

For non expert end users another kind of query interface is based on a predefined set of domain dependent queries. Every predefined query is composed of a description in natural language, describing a SPARQL query, with some free slots for parameters. Every parameter has a label, a type and a restriction on the valid values (e.g. a parameter can be filled only with instances of a specific class).

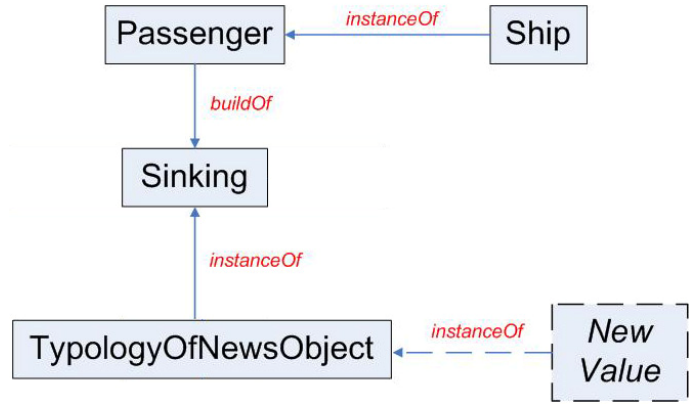

Figure 6. Graphical representation of the graph after the creation of the "New Value"

Figure 7. Creation of a new *instanceOf* for the literal value "New Value"

For this interface, users can select a query by its description, fill the query parameters and execute it. The results are presented as the results of the SPARQL query form. A future extension of this query mechanism may adapt the query with reference to the number of retrieved results.

VII. IMPLEMENTATION

The whole interaction is implemented in JavaScript and HTML5, often using the JavaScript to dynamically generate HTML5. The implementation was elaborated on Ajax-driven Web components [20, 21, 22] using relative sketch references of Web Application projects at Stanford University. JavaScript is used to handle the mouse events, create a pie chart object, and dynamically generate and update the HTML5 used to display graphically the data on the pie chart.

To draw the pie chart within the browser and to develop the accompanying form it was used HTML5 canvas.

Not essential to the implementation but also compatible with the code is the use of an input OWL file to load saved data and a PHP or other CGI script to process or save the chart information from the form.

*1) Javascript*

The described graphical interaction with the pie chart is done handling the mouse event towards JavaScript – eg. mouse move and mouse up, right click, single click, double click.

A JavaScript "pie chart object" is used to calculate its slices, and their angles, and to correlate the slices with the *Stress–strain* color representation for the elasticity in the *FD* ontology matching. JavaScript is also used to dynamically generate and update the HTML5 for the form displayed on the page.

The data used to reflect any changes within the graphical pie chart are regenerates through the DOM with the updated values from the form.

*2) HTML5*

The two main HTML5 components are the canvas and the form. The graphical chart is drawn on the HTML5 canvas [13, 14], which is cleared and redrawn every 50 ms to accurately reflect all changes made by the user. Once the first slice of the pie chart is created the form is generated and updated dynamically using HTML5 and JavaScript.

*3) OWL*

The user can pass an OWL file [23] that already contains pie chart formatted data with the correct tags of <name> and <percent> of the slices. Through the JavaScript *initPieChart()* function the pie chart is created. This default data are used to create the pie chart that user could then change it towards a graphical interaction with it. Without the OWL file the pie chart must be created using the form input having no slices.

*4) PHP*

The uri for a PHP [24] or CGI script can also be passed to *initPieChart()* function, telling the application what server side script to call to process the form once the user decides to save the chart inform.

## VIII. STRENGTHS AND WEAKNESES

*A. Strenghts*

The main strength is on the easy and intuitive graphical pie chart interface. The user is able to dynamically change the pie chart as he or she sees fit with simple mouse clicks and drags without having to fill spreadsheets or tables and to calculate previously the data values.

To reduce the overhead to the developer and to not requiring to the user to download additional software the current implementation does not use proprietary technology for the graphical pie chart development – such as Flash or Silverlight.

Data from the chart can easily be connected to a table while the pie chart event handlers can be done using mouse interactions. The canvas element will be a part of the HTML5 standard, so eventually; this interactive pie chart will be available to all browsers. The data processing in the browser was coded in JavaScript so it can be easy developed and encapsulated.

The call to *initPieChart()* function draws the interactive pie chart on the page developed on HTML5, using the JavaScript file included. Using a client side interaction, towards HTML5 and JavaScript, so there is no needs to fetch data while the user is interacting with the chart. The interaction can proceed smoothly not having to refresh the browser. However for a server interaction, parameters can be passed to *initPieChart()* function – specifying what script is used.

*B. Weaknesses*

Currently the main weakness of the implementation is on the slightly inefficient of the form component. To reflect any graphical changes made by the user, the script removes all previous data, and then dynamically regenerates the data for the correct chart representation. Even if the latency is negligible and so invisible to the user this is an inefficient implementation.

Further weakness is due to the difficult to save graphical information to be reused due to the use of HTML5 canvas [21] for the browser side implementation. Anyway all textual information can be saved through a CGI script (eg.: as the values of the slices, the names and the *stress–strain*).

Moreover using not proprietary technology it's obvious to have some browser compatibility issues. Because the implementation was developed using HTML5 and JavaScript it may not work on all browsers. It was tested and runs correctly on Firefox, Chrome and IE9, but it does not work properly on old version of IE that did not support for the canvas tag.

## IX. FUTURE DEVELOPMENT AND CONCLUSION

This paper introduced a Web–based system supporting the visualization, navigation and exploitation of folksonomies through the Web, and it also supports the updating and maintenance of this form of knowledge.

It is considered the concept of "*Folksodriven Structure Network*" to represent folksonomies based on the use of folksonomies and network theory. The network structure of Folksodriven tags – *Folksodriven Structure Network (FSN)* – was thought as a "Folksonomy tags suggestions" for the user on a dataset built on chosen websites.

The paper describes the functionalities offered by the system, that supports the exploitation of different existing semantic frameworks (i.e.: Jena, Sesame), and that presents an innovative HTML5–based web interface.

This work can be considered as an attempt to merge a semantic web context (Web 3.0) with a social aspect (Web 2.0). The system proposed can coordinate the accesses to a shared self constructed folksonomy KB.

Possible improvements can aimed at redesigning the user interface modules according to the model view controller pattern, further enhancing the usability of the web interface.

Others line of improvements:
*1) Multiple Charts*
The current implementation uses only a single pie chart. It would be useful to allow modification on multiple pie charts related to the same data.

*2) Value Assignment to Slices*
Currently to each slice in the chart the value assigned is a percentage. A possible improvement could be to allow the use of other kinds of value having those editable.

*3) Customizable Options*
A developer or user should be able to modify the pie chart according to their need (ie. how to semantically correlate the slice deletion respect other slices). Further improvement could be on more customizable features (ie. currently colors are used to represent the stress-strain but they should be used for other visualization purposes).

**Massimiliano Dal Mas** is an engineer at the Web Services division of the Telecom Italia Group, Italy. His interests include: user interfaces and visualization for information retrieval, automated Web interface evaluation and text analysis, empirical computational linguistics, text data mining, knowledge engineering and artificial intelligence. He received BA, MS degrees in Computer Science Engineering from the Politecnico di Milano, Italy. He won the thirteenth edition 2008 of the CEI Award for the best degree thesis with a dissertation on "Semantic technologies for industrial purposes" (Supervisor Prof. M. Colombetti).